\documentclass[12pt]{iopart}

\newcommand{\Dslash}{D \! \! \! \! /}

\newcommand{\half}{\mbox{\small{$\frac{1}{2}$}}} 
 
\newcommand{\Nf}{N_{\!f}} 
\newcommand{\NA}{N_{\!A}} 
\newcommand{\NF}{N_{\!F}} 
\newcommand{\Nda}{N^d_{\!A}} 
\newcommand{\Noda}{N^o_{\!A}} 
\newcommand{\MSbar}{\overline{\mbox{MS}}} 

\begin{document}

\title[RGE and the LCO method]{Renormalization group aspects of the local 
composite operator method\footnote{Talk presented at $5$th International
Conference on the Renormalization Group, Helsinki, Finland, 30th August - 3rd
September, 2005}} 

\author{R E Browne$^1$, D Dudal$^2$, J A Gracey$^1$, V E R Lemes$^3$, M S 
Sarandy$^4$, R F Sobreiro$^3$, S P Sorella$^3$ and H Verschelde$^2$}  

\address{$^1$ Theoretical Physics Division, Department of Mathematical 
Sciences, University of Liverpool, P.O. Box 147, Liverpool, L69 3BX, United
Kingdom \\
$^2$ Ghent University, Department of Mathematical Physics and Astronomy, 
Krijgslaan 281-S9, B-9000 Gent, Belgium \\ 
$^3$ Departamento de F\'{\i }sica Te\'{o}rica, Instituto de F\'{\i }sica, UERJ,
Universidade do Estado do Rio de Janeiro, Rua S{\~a}o Francisco Xavier 524, 
20550-013 Maracan{\~a}, Rio de Janeiro, Brazil \\
$^4$ Chemical Physics Theory Group, Department of Chemistry, University of
Toronto, 80 St George Street, Toronto, Ontario, M5S 3H6, Canada} 
\ead{jag@amtp.liv.ac.uk}
\begin{abstract}
We review the current status of the application of the local composite
operator technique to the condensation of dimension two operators in quantum
chromodynamics (QCD). We pay particular attention to the renormalization group
aspects of the formalism and the renormalization of QCD in various gauges. 
\end{abstract}

\vspace{-13cm}
\hspace{13cm}
{\bf LTH 692} 

\maketitle

\section{Introduction}

The quantum field theory underlying the strong force of nature is widely
accepted as being quantum chromodynamics (QCD) which is a non-abelian 
generalization of quantum electrodynamics. At large energy the constituent
fields, the quarks and gluons, behave as free particles which is a property
known as asymptotic freedom. In this case one performs calculations in QCD
based on a vacuum which is empty and which is known as the perturbative 
vacuum. However, it is accepted \cite{1,2,3,4} that the true vacuum of QCD is
more complicated and is not the perturbative one. An indication of this is that
in the true vacuum gauge invariant operators condense. Indeed the two operators
which receive wide attention are the operators $\alpha_S G^a_{\mu\nu} 
G^{a\,\mu\nu}$ and $\bar{\psi} \psi$, where $G^a_{\mu\nu}$ is the gluon field 
strength, $\psi$ is the quark field and $\alpha_S$ is the stong coupling
constant. Consequently, it is possible to incorporate the vacuum expectation 
values $\langle \alpha_S G^a_{\mu\nu} G^{a\,\mu\nu} \rangle$ and $\langle 
\bar{\psi} \psi \rangle$ into the operator product expansion in order to 
determine the effects they have in the measurements of physical quantities and 
QCD sum rules, \cite{5}. Indeed it is possible to extract numerical estimates 
for them. Whilst these operators are the main ones of interest, it has been 
pointed out more recently that additionally one can construct a dimension two 
operator in QCD which is gauge invariant, \cite{6,7,8}. However, this is also 
believed to condense giving rise to $O(1/Q^2)$ power corrections in the 
operator product expansion and other quantities, \cite{9,10,11}. Specifically 
the operator is
\begin{equation}  
\tilde{A}^2_\mu ~=~ \left[  
\stackrel{\mbox{min}}{\mbox{\begin{tiny}$\{U\}$\end{tiny}}} \int d^4x \, 
\left( A^U_\mu \right)^2 \right] {\cal V}^{-1}  
\label{nonlocop} 
\end{equation} 
where $U$ represents the set of all gauge configurations and $\tilde{A}^a_\mu$ 
is constructed in such a way that it is in fact gauge invariant. Consequently, 
unlike $G^a_{\mu\nu} G^{a\,\mu\nu}$ and $\bar{\psi} \psi$, the operator is 
non-local but can be written in terms of the usual gluon gauge field yielding a
power series in $g$ when evaluated explicitly. This dimension two operator has 
been the subject of intense study in recent years, mostly from the point of 
view of trying to estimate a value for its vacuum expectation value, 
\cite{12,13,14,15,16,17,18,19,20,21,22,23,24,25,26}. Further, the role a
non-vanishing vacuum expectation value of a dimension two operator had on the
estimate of glueball masses in the Coulomb gauge had been discussed earlier in 
\cite{27}. 

Having a non-zero vacuum expectation value for this operator has interesting
implications for trying to understand the properties of QCD and for
phenomenology. One area of study has been on the lattice where there appears
to be numerical evidence for $1/Q^2$ power corrections in a variety of
quantities \cite{9,10,11}. For example, an effective strong coupling constant, 
$\alpha^{\mbox{\footnotesize{eff}}}_S(Q^2)$, requires a $1/Q^2$ correction to 
correctly fit lattice data in the range $2$-$6$ GeV, \cite{28,29}. This 
necessitates a dimension two object on dimensional grounds. Another consequence 
is that such a dimension two condensate would imply that the gluon has an 
effective mass which is generated dynamically, \cite{13,14}. Estimates for the 
value of such a mass have been summarized in table $15$ of Field's article 
\cite{30}. These have been extracted from phenomenology where one includes a 
gluon with a mass in order to fit experimental data more accurately.

However, one of the main interests in understanding gluon mass is its
relationship to the confinement mechanism. Whilst there are various ideas about
what underlies this property of QCD and the strong force, the actual situation 
has not been determined yet. One point of view is that of abelian dominance 
\cite{31,32,33}. Essentially this is based on the premise that in the infrared 
the abelian sector of the gluon field dominates. It is then believed that the
infrared sector of QCD could be described by a dual superconductor, whereby a
monopole condensation would give rise to confinement via the dual Meissner
effect. Moreover, in the context of the generation of an effective gluon mass, 
one viewpoint is that in the infrared the gluons associated with the centre of
the colour group remain massless whilst the off-diagonal gluons gain a mass 
dynamically. Indeed there appears to be some preliminary lattice evidence for 
such a scenario, \cite{34,35}. To investigate such a hypothesis in QCD from a 
field theory point of view requires both a calculational technique to handle 
dimension two operator condensation as well as a way of focusing on the centre 
gluon fields. For the former, the local composite operator (LCO) method has 
been developed both for QCD, \cite{8}, and for models such as the two
dimensional Gross-Neveu model \cite{36,37}, where one has the exact mass gap to
justify the approach. To examine the differing nature of the gluon field, one 
can choose to fix in the maximal abelian gauge (MAG) where the gauge fixing 
differentiates between centre and off-diagonal gluons. In the main in this 
article we review the procedures and recent results in using the LCO method to 
study the consequences of the condensation of a dimension two operator in QCD 
in various gauges, concentrating on those aspects which relate to the 
renormalization group which underpins the technique.  

\section{Background} 
 
Before detailing the LCO approach it is worth recalling the background to the
problem of gluon mass in Yang-Mills theories. One early study was that of Curci
and Ferrari in \cite{38} where they constructed a Lagrangian with a gluon and 
ghost mass with a nonlinear gauge fixing. In particular the Lagrangian is
\begin{eqnarray} 
L &=& -~ \frac{1}{4} G_{\mu\nu}^A G^{A \, \mu\nu} ~-~ \frac{1}{2\alpha} 
(\partial^\mu A^A_\mu)^2 ~+~ \frac{m^2}{2} A_\mu^A A^{A \, \mu} \nonumber \\
&& +~ \partial_\mu \bar{c}^A \partial^\mu c^A ~-~ \alpha m^2 \bar{c}^A c^A ~-~ 
\frac{g}{2} f^{ABC} A^A_\mu \, \bar{c}^B \, {\stackrel \leftrightarrow 
{\partial^\mu} } \, c^C \nonumber \\
&& +~ \frac{\alpha g^2}{8} f^{EAB} f^{ECD} \bar{c}^A c^B \bar{c}^C c^D ~+~ 
i \bar{\psi} \Dslash \psi ~-~ m_q \bar{\psi} \psi 
\end{eqnarray}  
where $A^A_\mu$, $c^A$ and $\psi^{iI}$ are the respective gluon, ghost and
quark fields, $1$~$\leq$~$A$~$\leq$~$\NA$, $1$~$\leq$~$I$~$\leq$~$\NF$ and  
$1$~$\leq$~$i$~$\leq$~$\Nf$ with $\NA$ and $\NF$ the respective dimensions of
the adjoint and fundamental representations, $\Nf$ is the number of quarks,
$T^A$ are the generators of the colour group whose structure constants are 
$f^{ABC}$ and the field strength is given by  
$G^A_{\mu\nu}$~$=$~$\partial_\mu A^A_\nu$~$-$~$\partial_\nu A^A_\nu$~$-$
$g f^{ABC} A^B_\mu A^C_\nu$ where $g$ is the coupling constant. In the case 
when the gluon mass $m$ is zero, the Lagrangian is regarded as QCD fixed in the
Curci-Ferrari gauge. It gives rise to a different gluon-ghost interaction from 
that of the usual linear covariant gauge fixing. In addition there is a quartic
ghost interaction which does not invalidate the renormalizability of the 
Lagrangian. When $m$ is non-zero one has a mass for both the gluon and the 
ghost where the respective gluon and ghost propagators are 
\begin{equation}
-~ \frac{\delta^{AB}}{(k^2+m^2)} \left[ \eta^{\mu\nu} ~-~ 
\frac{(1-\alpha)k^\mu k^\nu}{(k^2+\alpha m^2)} \right] ~~~~ \mbox{and} ~~~~ 
\frac{\delta^{AB}}{(k^2+\alpha m^2)} ~.  
\end{equation} 
Whilst the Lagrangian is no longer invariant under gauge transformations, it is
in fact (on-shell) BRST invariant for non-zero $m$, \cite{38}. This latter 
property suggests it is a reasonable candidate for studying models with gluon 
mass. However, the initial interest in this model had to be tempered with the
realization that whilst one has BRST invariance the BRST charge is not
nilpotent since its square is propotional to $m^2$. Consequently one does not
have a unitary theory and negative norm states can be constructed to
demonstrate this \cite{39,40}. Aside from these limitations the Curci-Ferrari 
model has several important properties. One is that the presence of a mass for 
the gluon provides a natural infrared regulator in the theory. Indeed it has 
been renormalized explicitly at two loops, \cite{41,42}. Therefore, it could be
a useful tool in extracting renormalization constants where there are potential
infrared problems. More importantly though the resurgence of interest in this 
model rests in its relationship to other gauges. In the case where 
$\alpha$~$=$~$0$, the Curci-Ferrari gauge reduces to the usual Landau gauge,
\cite{38}. However, if one examines the off-diagonal sector of QCD fixed in the
maximal abelian gauge (MAG), it transpires that that sector is precisely QCD 
fixed in the Curci-Ferrari gauge, \cite{13}. Therefore, the Curci-Ferrari model
can be used as a laboratory for investigating the problem of abelian dominance 
in QCD and the dynamical generation of mass for the off-diagonal gluons. Whilst
the main disadvantage of the Curci-Ferrari model is the presence of a 
{\em classical} gluon mass leading to loss of unitarity, if a mass was 
generated dynamically by the condensation of a dimension two (BRST or gauge 
invariant) operator, then the unitarity issue may be circumvented. 

\section{LCO method} 

The LCO method is a procedure for including low dimension operators, such as
$\half A^A_\mu A^{A \, \mu}$, in the underlying quantum field theory and
determining its effective potential. In this way one can examine to what extent
the operator condenses by calculating whether the energy of the true vacuum 
when the operator is present is less than that of the true vacuum in its 
absence. For QCD it turns out that it leads to a modification of the Lagrangian
so that new interactions are introduced which lead to an effective gluon mass. 
Part of the justification in applying the LCO method to QCD in a variety of 
gauges, such as the Landau, Curci-Ferrari or MAG, lies in the treatment of the 
two dimensional $O(N)$ Gross-Neveu model. There the mass gap is known exactly 
and the LCO approach obtains values for the mass gap to a few percent for a 
large range of $N$, \cite{36,37}. 

We now summarise the application of the method in the case of QCD in the Landau
gauge. One of the advantages of this gauge is that the gauge invariant
non-local operator (\ref{nonlocop}) truncates to a single term local composite
operator $\half A^A_\mu A^{A \, \mu}$, \cite{8}. In this instance one couples 
the operator to a source $J$ yielding the energy functional $W[J]$ 
\begin{eqnarray} 
e^{-W[J]} &=& \int {\cal D} [A_\mu \psi \bar{\psi} c \bar{c}] \, \exp \left[ 
\int d^d x \left( L_{\mbox{\footnotesize{gf}}} - \frac{1}{2} Z_m J 
A_\mu^{A \, 2} + \frac{1}{2} ( \xi + \delta \xi ) J^2 \right) \right] ~.
\nonumber \\  
\end{eqnarray} 
From this, $W[J]$ satisfies a renormalization group equation 
\begin{equation} 
\left[ \mu \frac{\partial ~}{\partial \mu} + \beta(g) \frac{\partial ~}
{\partial g^2} - \gamma_m(g) \int J \frac{\partial ~}{\partial J} 
+ \eta(g,\xi) \frac{\partial ~}{\partial \xi} \right] W[J] ~=~ 0 
\end{equation} 
where $\gamma_m(g)$ is the anomalous dimension of the operator derived from
the corresponding renormalization constant $Z_m$ and $\mu$ is the 
renormalization scale introduced when one uses dimensional regularization in
$d$~$=$~$4$~$-$~$2\epsilon$ dimensions which is the regularization employed
here. To ensure renormalizability one requires the additional term quadratic in
$J$. This is because the vacuum energy in the presence of the operator is 
divergent with divergences proportional to $J^2$ appearing. The coefficient of 
$J^2$ is defined as $\xi$ where $\delta \xi$ is the counterterm and is at 
present not fixed, \cite{8}. However, one can define a renormalization group 
function for the infinities associated with the $J^2$ term which are encoded in
the related quantities $\eta(g,\xi)$ and $\delta(g)$ by 
\begin{eqnarray} 
\eta(g,\xi) &=& \mu \left. \frac{\partial \xi}{\partial \mu} \right| ~=~
2\gamma_m(g) \xi ~+~ \delta(g) \nonumber \\ 
\delta(g) &=& \left( 2\epsilon + 2\gamma_m(g) - \beta(g) 
\frac{\partial ~}{\partial g^2} \right) \delta \xi ~.  
\end{eqnarray} 
In order to have a homogeneous renormalization group equation for $W[J]$ the
as yet undetermined parameter $\xi$ must satisfy 
\begin{equation} 
\beta(g) \frac{d \xi}{d g^2} ~=~ 2 \gamma_m(g) \xi ~+~ \delta(g) 
\label{xidef}
\end{equation} 
whence 
\begin{equation} 
\left[ \mu \frac{\partial ~}{\partial \mu} + \beta(g) \frac{\partial ~}
{\partial g^2} - \gamma_m(g) \int J \frac{\partial ~}{\partial J} \right] 
W[J] ~=~ 0 ~. 
\label{wrge} 
\end{equation} 
Therefore solving (\ref{xidef}) will determine $\xi(g)$ once $\gamma_m(g)$ and 
$\delta(g)$ are known and this ensures that $\xi(g)$ runs as $g(\mu)$ runs. 
More importantly the homogeneity of (\ref{wrge}) ensures that one retains an 
energy interpretation so that an effective action and thence an effective 
potential can be constructed for the operator in question, \cite{8,36,37}. 

For practical calculations it would be more appropriate to have a functional
with a linear source. This can be achieved by a Hubbard-Stratonovich
transformation which introduces a scalar field $\sigma$ via  
\begin{equation} 
1 ~=~ \int {\cal D} \sigma \exp \left( \, - \int \left[ a_1 \sigma + 
a_2 A^{A \, 2}_\mu + a_3 J \right]^2 \right) 
\end{equation} 
where the coefficients $a_i$ are chosen appropriately to cancel the $J^2$
term. Consequently in the Landau gauge one obtains the renormalizable
Lagrangian for $\sigma$, and therefore the operator 
$\half A^A_\mu A^{A \, \mu}$ as
\begin{equation} 
L^\sigma ~=~ L_{\mbox{\footnotesize{gf}}} - \frac{\sigma^2}{2g^2 \xi(g) 
Z_\xi} + \frac{Z_m}{2 g \xi(g) Z_\xi} \sigma A^A_\mu A^{A \, \mu} 
- \frac{Z_m^2}{8\xi(g) Z_\xi} \left( A^A_\mu A^{A \, \mu} \right)^2 ~. 
\label{siglag} 
\end{equation} 
Once the expressions for $\gamma_m(g)$ and $\xi(g)$ are known then the 
effective potential can be constructed. Though for a two loop potential one
requires the renormalization group functions at three loops. 

\section{Three loop renormalization} 

As the LCO method relies on requiring explicit values of the renormalization
group functions at large loop order it is important to study the
renormalization of QCD in the context of the operator 
$\half A^A_\mu A^{A \, \mu}$ and in various gauges. For the Landau gauges
all the information to construct $\gamma_m(g)$ in fact is in place. This is due
to an observation from explicit calculations and the general formalism of 
algebraic renormalization which demonstrate that to all orders in perturbation
theory the anomalous dimension of $\half A^A_\mu A^{A \, \mu}$ is not 
independent, \cite{43}. More specifically 
\begin{equation} 
\gamma_m(g) ~=~ \gamma_A(g) ~+~ \gamma_c(g) 
\label{opstid}
\end{equation} 
in the Landau gauge. This curious property is not restricted to this gauge as 
in the MAG the anomalous dimension of the analogous dimension two operator,
based on off-diagonal fields, involves the anomalous dimensions of the 
diagonal gluon and diagonal ghost, \cite{26}. In the more general Curci-Ferrari
gauge we have observed a generalization of (\ref{opstid}) in an explicit three 
loop $\MSbar$ renormalization, \cite{44}, which is 
\begin{equation} 
\gamma_m(g) ~=~ \gamma_A(g) ~+~ \gamma_c(g) - 2\gamma_\alpha(g) 
\end{equation} 
where we note that unlike the linear covariant gauges the anomalous dimension 
corresponding to the renormalization of the gauge parameter,
$\gamma_\alpha(g)$, is non-zero. Unfortunately it has not been established
whether this latter relation remains valid beyond three loops. 

One issue which arises when one is working with the renormalization of
operators and this is the question of operator mixing. The BRST invariant mass
operator involves the two terms $\half A^A_\mu A^{A \, \mu}$ and 
$\bar{c}^A c^A$. In principle it could be the case that the combination 
${\cal O}$~$=$~$\half A^A_\mu A^{A \, \mu}$~$-$~$\alpha \bar{c}^A c^A$ does not 
renormalize multiplicatively. However, in linear covariant gauges it turns out 
that the mixing matrix is triangular, \cite{45}, but not in the Curci-Ferrari 
gauge. Indeed in \cite{46} the one loop mixing matrix for ${\cal O}_i$ was 
determined where ${\cal O}_1$~$=$~$\half A^A_\mu A^{A \, \mu}$ and 
${\cal O}_2$~$=$~$\bar{c}^A c^A$. We have extended that calculation to two
loops for potential future extensions of the operator product expansion
analysis of \cite{46}. If we set  
\begin{equation} 
{\cal O}_{o \, i} ~=~ Z_{ij} {\cal O}_j
\end{equation} 
where $Z_{ij}$ is the mixing matrix of renormalization constants. With 
\begin{equation} 
\gamma_{ij}(g) ~=~ \mu \frac{\partial ~}{\partial \mu} \ln Z_{ij}
\end{equation} 
then we have 
\begin{eqnarray} 
\gamma_{11} &=& \left( \frac{35}{12} + \frac{\alpha}{4} \right) C_A a ~+~ 
\left( \frac{449}{48} + \frac{11\alpha}{16} + \frac{3\alpha^2}{16} \right) 
C_A^2 a^2 ~+~ O(a^3) \nonumber \\ 
\gamma_{12} &=& - \frac{\alpha^2}{4} C_A a ~-~ 
\left( \frac{5\alpha^2}{16} + \frac{\alpha^3}{8} \right) C_A^2 a^2 ~+~ O(a^3) 
\nonumber \\ 
\gamma_{21} &=& \frac{\alpha}{2} C_A a ~-~ 
\left( \frac{11}{8} + \frac{\alpha}{4} \right) C_A^2 a^2 ~+~ O(a^3) 
\nonumber \\ 
\gamma_{22} &=& \left( \frac{3}{4} - \frac{\alpha}{4} \right) C_A a ~+~ 
\left( \frac{95}{48} + \frac{\alpha}{16} - \frac{\alpha^2}{8} \right) 
C_A^2 a^2 ~+~ O(a^3)  
\end{eqnarray} 
where $a$ $=$ $g^2/(16\pi^2)$, $T^A T^A$ $=$ $C_F I$, $f^{ACD} f^{BCD}$
$=$ $C_A \delta^{AB}$ and $\mbox{Tr} \left( T^A T^B \right)$ $=$ $T_F 
\delta^{AB}$. These results were obtained by renormalizing the operators in the
Curci-Ferrari model where there is a non-zero infrared regulating mass, by 
inserting them into gluon and ghost two-point functions. The Curci-Ferrari
model has the advantage that external momenta can be nullified without
introducing spurious infrared infinities as a consequence. It remains merely to
extract the infinities from the resultant vacuum bubbles. Not only did we 
reproduce the one loop matrix of Kondo, \cite{46}, but we obtained the result
that  
\begin{equation}
\gamma_m(g) ~=~ \gamma_{11}(g) ~-~ \alpha \gamma_{21}(g) 
\end{equation}
at two loops, thus verifying that ${\cal O}$ is multiplicatively 
renormalizable at this order. 

For three loop calculations the massive propagator approach is tedious and
we produced an equivalent method based on the {\sc Mincer} algorithm,  
\cite{47,48}, which is implemented in the symbolic manipulation language 
{\sc Form}, \cite{49}. For example, one can determine $\delta \xi$ by treating 
the term $\half J A^A_\mu A^{A \, \mu}$ of (\ref{siglag}) as an interaction and
computing the divergence structure of the $J$ two-point function with massless
internal fields but not internal $J$ propagators, \cite{24}. The explicit 
Feynman diagrams are generated automatically with the {\sc Qgraf} package,
\cite{50}. The {\sc Mincer} algorithm was especially appropriate for the three 
loop renormalization of QCD in the MAG, \cite{51}, which is necessary for the 
construction of the two loop effective potential for the analogous dimension 
two BRST invariant operator. Unlike the linear covariant gauges the full three 
loop renormalization was determined only recently, \cite{51}. Moreover, it was 
a significantly large computation requiring the evaluation of $37322$ Feynman
diagrams compared with of the order of $1000$ for a linear covariant gauge
three loop renormalization.

Briefly, the MAG involves the decomposition of the gauge field $A^A_\mu$
into diagonal and off-diagonal sectors 
\begin{equation} 
A^A_\mu T^A ~=~ A^a_\mu T^a ~ + ~ A^i_\mu T^i  
\end{equation}
where $1$~$\leq$~$a$~$\leq$~$\Noda$ and $1$~$\leq$~$i$~$\leq$~$\Nda$ and $\Nda$
is the dimension of the centre of the colour group and $\Noda$ is the dimension
of the remainder with $\Nda$~$+$~$\Noda$~$=$~$\NA$. Notationally we will 
reserve $i$, $j$, $k$ and $l$ for indices on objects which lie in the centre of
the group and the remaining lower case Roman letters for off-diagonal objects. 
Consequently, the MAG gauge fixing term is, \cite{26}, 
\begin{equation} 
L^{\mbox{\footnotesize{MAG}}}_{\mbox{\footnotesize{gf}}} ~=~ \delta 
\bar{\delta} \left[ \half A_\mu^a A^{a \, \mu} ~+~ \half \alpha \bar{c}^a 
c^a \right] ~+~ \delta \left[ \bar{c}^i \partial^\mu A_\mu^i \right] 
\end{equation} 
where $\delta$ and $\bar{\delta}$ are the BRST and anti-BRST transformations.
The remaining gauge freedom associated with the diagonal gluons is fixed by
using a Landau gauge. Further, the analogous mass operator to ${\cal O}$ is  
\begin{equation} 
{\cal O}^{\mbox{\footnotesize{MAG}}} ~=~ \half A^a_\mu A^{a \, \mu} ~-~ 
\alpha \bar{c}^a c^a ~. 
\end{equation} 
To renormalize the resultant Lagrangian 
\begin{eqnarray}
L^{\mbox{\footnotesize{MAG}}}_{\mbox{\footnotesize{gf}}} &=& -~ 
\frac{1}{2\alpha} \left( \partial^\mu A^a_\mu \right)^2 
- \frac{1}{2\bar{\alpha}} \left( \partial^\mu A^i_\mu \right)^2 
+ \bar{c}^a \partial^\mu \partial_\mu c^a
+ \bar{c}^i \partial^\mu \partial_\mu c^i \nonumber \\
&& +~ g \left[ f^{abk} A^a_\mu \bar{c}^k \partial^\mu c^b 
- f^{abc} A^a_\mu \bar{c}^b \partial^\mu c^c \right. \nonumber \\
&& \left. ~~~~~~~ 
-~ \frac{1}{\alpha} f^{abk} \partial^\mu A^a_\mu A^b_\nu A^{k \, \nu} 
- f^{abk} \partial^\mu A^a_\mu c^b \bar{c}^k \right. \nonumber \\
&& \left. ~~~~~~~ 
-~ \frac{1}{2} f^{abc} \partial^\mu A^a_\mu \bar{c}^b c^c 
- 2 f^{abk} A^k_\mu \bar{c}^a \partial^\mu \bar{c}^b 
- f^{abk} \partial^\mu A^k_\mu \bar{c}^b c^c \right] \nonumber \\  
&& +~ g^2 \left[ f_d^{acbd} A^a_\mu A^{b \, \mu} \bar{c}^c c^d  
- \frac{1}{2\alpha} f_o^{akbl} A^a_\mu A^{b \, \mu} A^k_\nu
A^{l \, \nu} \right.  
\nonumber \\
&& \left. ~~~~~~~~  
+~ f_o^{adcj} A^a_\mu A^{j \, \mu} \bar{c}^c c^d 
- \frac{1}{2} f_o^{ajcd} A^a_\mu A^{j \, \mu} \bar{c}^c c^d \right.  
\nonumber \\
&& \left. ~~~~~~~~  
+~ f_o^{ajcl} A^a_\mu A^{j \, \mu} \bar{c}^c c^l  
+ f_o^{alcj} A^a_\mu A^{j \, \mu} \bar{c}^c c^l  
- f_o^{cjdi} A^i_\mu A^{j \, \mu} \bar{c}^c c^d \right.  
\nonumber \\
&& \left. ~~~~~~~~  
-~ \frac{\alpha}{4} f_d^{abcd} \bar{c}^a \bar{c}^b c^c c^d  
- \frac{\alpha}{8} f_o^{abcd} \bar{c}^a \bar{c}^b c^c c^d \right.  
\nonumber \\
&& \left. ~~~~~~~~  
+~ \frac{\alpha}{8} f_o^{acbd} \bar{c}^a \bar{c}^b c^c c^d 
- \frac{\alpha}{4} f_o^{abcl} \bar{c}^a \bar{c}^b c^c c^l  
\right. \nonumber \\
&& \left. ~~~~~~~~  
+~ \frac{\alpha}{4} f_o^{acbl} \bar{c}^a \bar{c}^b c^c c^l  
- \frac{\alpha}{4} f_o^{albc} \bar{c}^a \bar{c}^b c^c c^l 
+ \frac{\alpha}{2} f_o^{akbl} \bar{c}^a \bar{c}^b c^k c^l \right] \nonumber  
\end{eqnarray}   
where
\begin{equation}
f_d^{ABCD} ~=~ f^{iAB} f^{iCD} ~~,~~ 
f_o^{ABCD} ~=~ f^{eAB} f^{eCD} 
\end{equation}
one introduces renormalization constants via, \cite{26,52,53,54,55,56},   
\begin{eqnarray} 
A^{a \, \mu}_{\mbox{\footnotesize{o}}} &=& \sqrt{Z_A} \, A^{a \, \mu} ~~,~~ 
A^{i \, \mu}_{\mbox{\footnotesize{o}}} ~=~ \sqrt{Z_{A^i}} \, A^{i \, \mu} 
\nonumber \\ 
c^a_{\mbox{\footnotesize{o}}} &=& \sqrt{Z_c} \, c^a ~~,~~ 
\bar{c}^a_{\mbox{\footnotesize{o}}} ~=~ \sqrt{Z_c} \, \bar{c}^a \nonumber \\
c^i_{\mbox{\footnotesize{o}}} &=& \sqrt{Z_{c^i}} \, c^i ~~,~~ 
\bar{c}^i_{\mbox{\footnotesize{o}}} ~=~ \frac{\bar{c}^i}{\sqrt{Z_{c^i}}} ~~,~~ 
\psi_{\mbox{\footnotesize{o}}} ~=~ \sqrt{Z_\psi} \psi ~, \nonumber \\  
g_{\mbox{\footnotesize{o}}} &=& \mu^\epsilon Z_g \, g ~~,~~ 
\alpha_{\mbox{\footnotesize{o}}} ~=~ Z^{-1}_\alpha Z_A \, 
\alpha ~,~  
\bar{\alpha}_{\mbox{\footnotesize{o}}} ~=~ Z^{-1}_{\alpha^i} Z_{A^i} \, 
\bar{\alpha} ~.  
\end{eqnarray} 
However, it is crucial to note that this choice is determined by the 
application of the algebraic renormalization method, \cite{26}. This shows, for
example, that the diagonal ghost two-point function is finite to all orders and
implies that its anomalous dimension must be deduced from another Green's 
function such as the $A^a_\mu \bar{c}^i c^b$ vertex. Also, the diagonal gluon 
anomalous dimension is not independent since its associated renormalization 
constant is equivalent to that for the coupling constant, \cite{26}. A similar 
feature occurs in the background field gauge, \cite{57,58,59,60}. Whilst the 
application of the {\sc Mincer} algorithm is straightforward to extract all the
necessary renormalization constants, the bulk of the work lies in symbolically 
implementing the underlying group theory relations founded upon the elementary 
equations
\begin{equation}
f^{ijk} ~=~ f^{ajk} ~=~ 0 ~~,~~ f^{abk} ~\neq~ 0 ~~,~~ f^{abc} ~\neq~ 0 ~.
\end{equation}
Consequently one obtains representative anomalous dimensions of the following
form in the $\MSbar$ scheme  
\begin{eqnarray} 
\gamma_{c^i}(a) &=& \frac{1}{4 \Noda} \left[ \Noda \left( ( - \alpha - 3 ) C_A
\right) + \Nda \left( ( - 2 \alpha - 6 ) C_A \right) \right] a \nonumber \\
&& +~ \frac{1}{96 {\Noda}^2} \left[ {\Noda}^2 \left( ( -~ 6 \alpha^2 
- 66 \alpha - 190 ) C_A^2 + 80 C_A T_F \Nf \right) \right. \nonumber \\
&& \left. ~+~ \Noda \Nda \left( ( -~ 54 \alpha^2 - 354 \alpha
- 323 ) C_A^2 + 160 C_A T_F \Nf \right) \right. \nonumber \\
&& \left. ~+~ {\Nda}^2 \left( ( -~ 60 \alpha^2 - 372 \alpha 
+ 510 ) C_A^2 \right) \right] a^2 \nonumber \\
&& +~ \frac{1}{6912 {\Noda}^3} \left[ {\Noda}^3 ( ( -~ 162 \alpha^3 
- 2727 \alpha^2 - 2592 \zeta_3 \alpha - 18036 \alpha 
\right. \nonumber \\
&& \left. ~ 
-~ 1944 \zeta_3 - 63268 ) C_A^3 
+ ( 6912 \alpha + 62208 \zeta_3 + 6208 ) C_A^2 T_F \Nf 
\right. \nonumber \\
&& \left. ~
+~ ( -~ 82944 \zeta_3 + 77760 ) C_A C_F T_F \Nf 
+ 8960 C_A T_F^2 \Nf^2 )
\right. \nonumber \\
&& \left. ~
+~ {\Noda}^2 \Nda ( ( -~ 2754 \alpha^3 + 648 \zeta_3 \alpha^2 
- 28917 \alpha^2 - 4212 \zeta_3 \alpha 
\right. \nonumber \\
&& \left. ~
-~ 69309 \alpha 
+ 37260 \zeta_3 - 64544 ) C_A^3 
\right. \nonumber \\
&& \left. ~
+~ ( 25488 \alpha + 103680 \zeta_3 - 13072 ) C_A^2 T_F \Nf 
\right. \nonumber \\
&& \left. ~ 
+~ ( -~ 165888 \zeta_3 + 155520 ) C_A C_F T_F \Nf 
+ 17920 C_A T_F^2 \Nf^2 )
\right. \nonumber \\
&& \left. ~
+~ \Noda {\Nda}^2 ( ( -~ 7884 \alpha^3 + 22680 \zeta_3 \alpha^2
- 84564 \alpha^2 + 97524 \zeta_3 \alpha 
\right. \nonumber \\
&& \left. ~
-~ 47142 \alpha + 433836 \zeta_3 - 56430 ) C_A^3 
\right. \nonumber \\
&& \left. ~
+~ ( 25056 \alpha - 124416 \zeta_3 - 18144 ) C_A^2 T_F \Nf )
\right. \nonumber \\
&& \left. ~
+~ {\Nda}^3 ( ( -~ 6480 \alpha^3 + 34992 \zeta_3 \alpha^2 
- 70092 \alpha^2 + 8424 \zeta_3 \alpha 
\right. \nonumber \\
&& \left. ~ 
+~ 114912 \alpha + 77112 \zeta_3 - 161028 ) C_A^3 ) \right] a^3 ~+~ O(a^4) 
\end{eqnarray} 
and 
\begin{eqnarray} 
\gamma_{\cal O}(a) &=& \frac{1}{12 \Noda} \left[ \Noda \left( ( -~ 3 \alpha 
+ 35 ) C_A - 16 T_f \Nf \right) + \Nda \left( ( -~ 6 \alpha - 18 ) C_A \right) 
\right] a \nonumber \\
&& +~ \frac{1}{96 {\Noda}^2} \left[ {\Noda}^2 \left(  ( -~ 6 \alpha^2 
- 66 \alpha + 898 ) C_A^2 - 560 C_A T_f \Nf 
\right. \right.  \nonumber \\
&& \left. \left. ~-~ 384 C_F T_f \Nf \right) 
\right. \nonumber \\
&& \left. 
+~ \Noda \Nda \left(  ( -~ 54 \alpha^2 - 354 \alpha 
- 323 ) C_A^2 + 160 C_A T_f \Nf \right) \right. \nonumber \\
&& \left. ~+~ {\Nda}^2 \left(  ( -~ 60 \alpha^2 - 372 \alpha 
+ 510 ) C_A^2 \right) \right] a^2 \nonumber \\
&& +~ \frac{1}{6912 {\Noda}^3} \left[ {\Noda}^3 ( ( -~ 162 \alpha^3 
- 2727 \alpha^2 - 2592 \zeta_3 \alpha - 18036 \alpha 
\right. \nonumber \\
&& \left. ~
-~ 1944 \zeta_3 + 302428 ) C_A^3 
\right. \nonumber \\
&& \left. ~
+~ ( 6912 \alpha + 62208 \zeta_3 - 356032 ) C_A^2 T_F \Nf 
\right. \nonumber \\
&& \left. ~
+~ ( -~ 82944 \zeta_3 
- 79680 ) C_A C_F T_F \Nf + 49408 C_A T_F^2 \Nf^2 
\right. \nonumber \\
&& \left. ~
+~ 13824 C_F^2 T_F \Nf 
+ 33792 C_F T_F^2 \Nf^2 )
\right. \nonumber \\
&& \left. ~
+~ {\Noda}^2 {\Nda} ( ( -~ 2754 \alpha^3 + 648 \alpha^2 \zeta_3
- 28917 \alpha^2 
\right. \nonumber \\
&& \left. ~
-~ 4212 \alpha \zeta_3 
- 69309 \alpha + 37260 \zeta_3 - 64544 ) C_A^3
\right. \nonumber \\
&& \left. ~
+~ ( 25488 \alpha + 103680 \zeta_3 
- 13072 ) C_A^2 T_F \Nf 
\right. \nonumber \\
&& \left. ~
+~ ( -~ 165888 \zeta_3 + 155520 ) C_A C_F T_F \Nf 
+ 17920 C_A T_F^2 \Nf^2 ) 
\right. \nonumber \\
&& \left. ~
+~ {\Noda} {\Nda}^2 (  ( -~ 7884 \alpha^3 + 22680 \alpha^2 \zeta_3
- 84564 \alpha^2 + 97524 \alpha \zeta_3 
\right. \nonumber \\
&& \left. ~
-~ 47142 \alpha 
+ 433836 \zeta_3 - 56430 ) C_A^3
+ ( 25056 \alpha - 124416 \zeta_3 
\right. \nonumber \\
&& \left. ~
-~ 18144 ) C_A^2 T_F \Nf ) 
+ {\Nda}^3 (  ( -~ 6480 \alpha^3 + 34992 \alpha^2 \zeta_3 
- 70092 \alpha^2 
\right. \nonumber \\
&& \left. ~
+~ 8424 \alpha \zeta_3 
+ 114912 \alpha 
+ 77112 \zeta_3 - 161028 ) C_A^3) \right] a^3 ~+~ O(a^4) \nonumber \\  
\end{eqnarray} 
for the MAG mass operator where $\zeta_n$ is the Riemann zeta function, 
\cite{51}. In addition the three loop $\beta$-function correctly emerges from 
the diagonal gluon two-point function which is a strong check on the 
programming and computation since not only must it be independent of the gauge 
parameter $\alpha$ but also of the sector dimensions $\Nda$ and $\Noda$. 
Another useful check on this and the anomalous dimensions was the fact that 
the known Curci-Ferrari gauge anomalous dimensions, \cite{41,42,44}, emerge in 
the limit $\Nda/\Noda$~$\rightarrow$~$0$. This is consistent with the relation 
of the Curci-Ferrari gauge to the MAG, \cite{13}. 

\section{Results}

Having detailed the renormalization group aspects underlying the LCO formalism
we now briefly summarize recent results of determining estimates for the gluon
mass in various gauges, \cite{8,24,26}. First, for the Landau gauge the 
effective potential for $\sigma$ is, \cite{8,24},  
\begin{eqnarray}
V(\sigma) &=& 
\frac{9N_A}{2} \lambda_1 \sigma^{\prime \, 2} \nonumber \\
&& +~ \left[ \frac{3}{64} \ln \left( \frac{g \sigma^\prime}{\bar{\mu}^2} 
\right)
- C_A \left(
\frac{351}{8} C_F \lambda_1 \lambda_2
- \frac{351}{16} C_F \lambda_1 \lambda_3
\right. \right. \nonumber \\
&& \left. \left. ~~~~~ 
+~ \frac{249}{128} \lambda_2
- \frac{27}{64} \lambda_3
\right)
+ C_A^2 \left(
-~ \frac{81}{16} \lambda_1 \lambda_2
+ \frac{81}{32} \lambda_1 \lambda_3
\right)
\right. \nonumber \\
&& \left. ~~~~~ 
+ \left(
-~ \frac{13}{128}
- \frac{207}{32} C_F \lambda_2
+ \frac{117}{32} C_F \lambda_3
\right)
\right] \frac{g^2 N_A \sigma^{\prime \, 2}}{\pi^2} ~+~ O(g^4) \nonumber  
\end{eqnarray} 
where space has restricted us to the one loop expression and 
$\lambda_1$ $=$ $[13 C_A-8 T_F \Nf]^{-1}$, 
$\lambda_2$~$=$~$[35 C_A-16 T_F \Nf]^{-1}$, 
$\lambda_3$~$=$~$[19 C_A-8 T_F \Nf]^{-1}$ and
$\sigma$~$=$~$\frac{9 N_A}{(13 C_A - 8 T_F \Nf)} \sigma^\prime$. Examining
the solution to $V^\prime(\sigma)$ $=$ $0$ there are two possibilities which
are $\langle \sigma \rangle$ $=$ $0$ or $\langle \sigma \rangle$ $\neq$ $0$.
For the former this is the original classical vacuum but the latter corresponds
to a new vacuum which has an energy lower than the former. Thus in the
presence of the $\half A^A_\mu A^{A \, \mu}$ operator the effective potential
produces a new vacuum which is stable unlike the now unstable (perturbative)
classical vacuum. Moreover, boundedness of the potential requires that
$[13 C_A - 8 T_F \NF]$ needs to be positive, \cite{24}. Interestingly this 
corresponds to the Landau gauge one loop gluon anomalous dimension which has 
been suggested as part of the necessary criterion underlying confinement when
that problem is considered from a renormalization group perspective, 
\cite{61,62}. Consequently if one defines 
$m^2_{\mbox{\footnotesize{eff}}}$~$=$~$\sigma/(g\xi(g))$ as an effective gluon 
mass then for $SU(3)$ Yang-Mills $m_{\mbox{\footnotesize{eff}}}$~$=$~$2.13 
\Lambda_{\mbox{\footnotesize{$\MSbar$}}}$ from the two loop potential 
\cite{8,24}. This is within $2\%$ of the one loop estimate indicating a degree 
of stability in the approach. As an alternative one can compute the gluon pole 
mass by first redefining $\sigma^\prime$ in terms of the pole mass and 
demanding the alternative condition, \cite{63,64}, 
\begin{equation}
\frac{d V(m_{\mbox{\footnotesize{pole)}}}}
{d m_{\mbox{\footnotesize{pole}}}} ~=~ 0 ~. 
\end{equation}  
Interestingly at one loop this produces a Yang-Mills mass which is independent 
of the renormalization scale, \cite{63}. Though at two loops, like the 
effective mass of \cite{8}, the pole mass derived from the effective potential
is scale dependent. 

For the MAG the analysis is not fully complete as only the one loop potential
for $SU(2)$ has been determined, \cite{26}. However, the situation there is 
encouraging in that for pure Yang-Mills a mass is generated for the 
off-diagonal gluons which is $m_{\mbox{\footnotesize{eff}}}$~$=$~$2.25 
\Lambda_{\mbox{\footnotesize{$\MSbar$}}}$. This is not dissimilar to the
Landau gauge $SU(2)$ estimate of $m_{\mbox{\footnotesize{eff}}}$~$=$~$2.03 
\Lambda_{\mbox{\footnotesize{$\MSbar$}}}$. In addition the off-diagonal
ghost and diagonal gluon remain massless. The appearance of the potential
diagonal gluon mass operator, $\half A^i_\mu A^{i \, \mu}$, in the LCO action
used for the MAG, \cite{26}, is excluded by the diagonal $U(1)$ Ward identity
deriving from the algebraic renormalization analysis, \cite{26}. We are unable 
to prove the renormalizability of the action supplemented with a mass term like 
$\half {\cal J} A^i_\mu A^{i \, \mu}$. Indeed overall this mass generation
scenario appears to be consistent with $SU(2)$ lattice studies in the maximal 
abelian gauge, \cite{34,35,65}. 

\section{Conclusions}

We conclude with various observations. First, we have given an overview of the
current status of the application of the local composite operator method to the
condensation of a renormalizable dimension two operator in QCD in various 
gauges, concentrating on the underlying renormalization group aspects. One main
feature is the construction of a two loop effective potential for the operator
which requires knowledge of the three loop anomalous dimensions of QCD. Whilst
these are known for linear covariant gauges, to examine the abelian dominance
hypothesis in the infrared, the more appropriate maximal abelian gauge needs to
be used. This has required the full three loop renormalization of QCD in the
MAG, which is a significantly larger computation from the point of view of the 
number of Feynman diagrams to be evaluated. Moreover, it opens up the 
possibility of examining the generation of a mass for the off-diagonal gluon at
the two loop level and for gauge groups other than $SU(2)$. Whilst this may 
seem to be a feature of this gauge, the issue of whether one can access
abelian dominance in a covariant gauge, where the properties of the centre of
the group are not explicit in the fields one uses, has recently been studied
using the LCO formalism, \cite{66}. In particular the presence of ghost 
condensates in $SU(2)$ appears to be central in the dynamical generation of
a mass for the off diagonal gluons which is different from that of the diagonal
gluons. Indeed there would appear to be evidence from a recent lattice study to 
support this point of view, \cite{67}.   

\ack 
We gratefully acknowledge the financial support which was provided by the 
Conselho Nacional de Desenvolvimento Cient\'{i}fico e Tecnol\'{o}gico 
(CNPq-Brazil), the Faperj, Funda{\c{c}}{\~{a}}o de Amparo {\`{a}} Pesquisa do 
Estado do Rio de Janeiro, the SR2-UERJ, the Coordena{\c{c}}{\~{a}}o de 
Aperfei{\c{c}}oamento de Pessoal de N{\'\i}vel Superior (CAPES), PPARC for a
research studentship and the \emph{Special Research Fund} of Ghent University.

\end{document}